\documentclass[a4paper,twocolumn,nopdfoutputerror]{quantumarticle}
\ifdefined\pdfoutput\pdfoutput=1\fi
\usepackage[utf8]{inputenc}
\usepackage[english]{babel}
\usepackage[T1]{fontenc}
\usepackage{amsmath}
\usepackage{amssymb}
\usepackage{graphicx}
\usepackage{hyperref}

\usepackage[numbers]{natbib}
\usepackage{mathrsfs}          
\usepackage{booktabs}          
\usepackage{algpseudocode}     

\newcounter{algorithm}
\renewcommand{\thealgorithm}{\arabic{algorithm}}
\newenvironment{algorithm}[1][]{%
  \refstepcounter{algorithm}%
  \par\vspace{8pt}\noindent%
  \begin{minipage}{\linewidth}%
  \centering\bfseries Algorithm~\thealgorithm%
  \ifx&#1&\else:~#1\fi%
  \par\vspace{4pt}%
}{%
  \end{minipage}%
  \par\vspace{8pt}%
}
\usepackage{listings}          
\usepackage{anyfontsize}       
\usepackage{subfig}            

\graphicspath{{./images/}}

\begin{document}

\title{Physics-Informed Quantum Machine Learning with Hard Constraint Embedding for Nonlinear Differential Equations of the First Order}

\author[1]{Mengke Xu}
\email{xmk22@cjlu.edu.cn}
\author[2]{Xi Li}
\author[1]{Xiao Chen}
\author[1]{Xunan Wang}
\author[1]{Wanli Huo}
\author[4]{Long Ma}
\author[3]{Weiqi Yan}

\affil[1]{Zhejiang-New Zealand Joint Laboratory on Vision-Based Intelligent Metrology, School of Information Engineering, China Jiliang University, Hangzhou, China}
\affil[2]{School of Software, Henan University, Zhengzhou, Henan, China}
\affil[3]{Zhejiang-New Zealand Joint Laboratory on Vision-Based Intelligent Metrology, Computer Mathematical Sciences, Auckland University of Technology, Auckland, New Zealand}
\affil[4]{Alipay Digital Services Technology, Hangzhou, Zhejiang, China}

\maketitle

\begin{abstract}
Quantum algorithms based on linear-system approaches for solving differential equations demand qubit and precision resources beyond near-term capabilities. To address these challenges, this work proposes a physics-informed quantum machine learning (PIQML) framework with hard constraint embedding, specifically designed for NISQ era. Within this framework, parameterized quantum circuits serve as machine learning models, where the input variable is encoded into a high-dimensional feature space via a Fourier feature map. Subsequently, to eliminate approximation errors in critical physical conditions, the solution is constructed through a rigorously designed function mapper that analytically enforces initial conditions as hard constraints. Crucially, we compute derivatives with respect to the input variable using the parameter-shift rule---a quantum native gradient evaluation technique that avoids classical discretization. Unlike generic loss functions that target abstract data patterns, our loss function focuses on the differential equation residual and reference data. This design ensures that the trained model not only approximates the data but also intrinsically satisfies the physical constraint expressed by the DE itself. Our method is validated on several differential equations, including highly oscillatory ones, demonstrating its capability to tackle challenging nonlinear dynamics. Results demonstrate that our quantum model successfully learns the solution, showing close agreement with a high-precision classical numerical benchmark.

\vspace{5pt}
\end{abstract}

\section{Introduction}\label{sec1}

Differential equations constitute a foundational tool for scientific modeling, with applications spanning mechanical dynamics, financial systems, and epidemiology. The original motivation for developing quantum computers was to efficiently simulate quantum mechanical systems, as governed by the Schr\"{o}dinger equation---a task known to be exponentially challenging for classical computers in general cases~\cite{nielsen2010quantum}. This natural advantage in simulating quantum physics suggests a broader potential for quantum computers to solve general classes of differential equations (DEs) more efficiently.

Classical numerical methods for solving DEs, such as the finite difference method (FDM)~\cite{najmuddin2022study} and spectral methods (SM)~\cite{kurdi2008spectral}, typically discretize the equations into large systems of linear equations. Early quantum algorithms adopted a similar linear-algebraic approach, most notably through the Harrow--Hassidim--Lloyd (HHL) algorithm, which offered exponential speedups in theory for linear and constant-coefficient DEs~\cite{cao2013quantum,berry2014high,berry2017quantum,childs2020quantum}. However, these methods generally demand deep, fault-tolerant quantum circuits and can suffer from the curse of dimensionality, rendering them impractical for the current era of Noisy Intermediate-Scale Quantum (NISQ) devices~\cite{preskill2018quantum}.

The NISQ era has instead stimulated the development of hybrid quantum--classical algorithms, which combine parameterized quantum circuits (PQCs) with classical optimization. In such frameworks, PQCs are employed as tunable models whose parameters are iteratively adjusted by a classical optimizer to minimize a problem-specific cost function~\cite{benedetti2019parameterized}. This approach has proven effective across several domains, including quantum chemistry with the Variational Quantum Algorithm (VQA)~\cite{mcclean2016theory,kandala2017hardware,wang2018quantum,bravo2023variational,allcock2020quantum}, combinatorial optimization with the Quantum Approximate Optimization Algorithm (QAOA)~\cite{farhi2014quantum}, and more recently, supervised and generative machine learning tasks~\cite{benedetti2019parameterized}. By delegating the bulk of the parameter search to classical routines, these variational algorithms significantly relax the demands on quantum coherence time and gate fidelity, making them among the most promising strategies for attaining practical quantum advantage in the near term~\cite{liu2025variational,sato2021variational,leong2022variational}.

In parallel, the field of scientific machine learning has advanced considerably, notably with the emergence of Physics-Informed Neural Networks (PINNs)~\cite{raissi2019physics,cai2021physics}. PINNs integrate neural networks with the governing physical laws expressed as differential equations, thereby ensuring that the learned solutions are consistent with the underlying physics.

Building on the success of PINNs and motivated by the quest for quantum-enhanced simulation, the nascent field of Physics-Informed Quantum Machine Learning (PIQML) has emerged~\cite{kyriienko2021solving,setty2025self}. This area has attracted growing attention, with several recent studies exploring quantum physics-informed neural networks from different perspectives~\cite{trahan2024quantum,sedykh2024hybrid,siegl2025solving,markidis2022physics,knudsen2020solving,panichi2026quantum}. PIQML aims to harness the expressive power of parameterized quantum circuits as compact function approximators, potentially offering advantages in capacity or trainability for certain problems. However, existing PIQML methods for solving differential equations---especially nonlinear ones---face notable limitations. A key issue is the soft constraint treatment of initial and boundary conditions (I/BCs), wherein these conditions are incorporated as penalty terms in the loss function. This approach often leads to delicate trade-offs during optimization, poor convergence, and solutions that do not strictly satisfy the physical constraints~\cite{paine2023physics}.

To overcome these shortcomings, we propose a physics-informed quantum machine learning framework with hard constraint embedding, specifically designed for NISQ devices. Our work makes the following contributions:

\textbf{Hard Constraint Embedding via Analytical Mappers:} We design a quantum-circuit-based function mapper that analytically and exactly enforces initial and boundary conditions by construction. This transforms a constrained optimization into an unconstrained one, guarantees strict adherence to the physical constraints, and eliminates the balancing act inherent in penalty-based methods.

\textbf{Quantum Native Derivative Computation:} We eschew classical discretization of derivatives. Instead, we leverage the parameter-shift rule---a gradient estimation technique native to PQCs---to compute derivatives directly within the quantum circuit, enabling a fully quantum-native gradient evaluation for the differential equation residual.

\textbf{Integrated PIQML Loss Formulation:} Within our physics-informed quantum learning paradigm, we construct a streamlined loss function that balances the numerical fit to reference data with the physical fidelity enforced by the differential equation residual.

\textbf{NISQ-Compatible Architecture:} The entire framework is built around shallow, hardware-efficient PQCs, avoiding the need for deep circuits or fault-tolerant quantum computation.

The remainder of this paper is organized as follows. In Section~\ref{sec2}, we describe the methodology, including feature maps (Section~\ref{subsec2}), the variational ansatz, the hard constraint embedding mechanism, the derivative computation via the parameter-shift rule, and the physics-informed loss function. In Section~\ref{sec3}, we present numerical results on several benchmark problems, including a first-order linear ODE, a first-order polynomial ODE, a parametrized damped oscillator equation, and a nonlinear oscillatory equation. In Section~\ref{sec4}, we conclude with a discussion of the findings and future research directions.

\section{Methods}\label{sec2}

Building upon the core framework outlined in Section~\ref{sec1}, this section details the key methodological components of our proposed physics-informed quantum machine learning with hard constraint embedding for nonlinear differential equations. We begin by describing our feature encoding strategies.

\subsection{Feature maps}

To capture complex solution structures in variational quantum circuits, we apply a targeted feature mapping strategy. For robust feature encoding, we use a Chebyshev feature map, which encodes the input variable $x$ using a basis of orthogonal polynomials known for their numerical stability and efficiency~\cite{trefethen2019approximation}. For solutions dominated by high-frequency oscillations, we supplement it with a Fourier feature map. This map explicitly projects the input $x$ into a high-dimensional frequency space spanned by sinusoidal basis functions, to capture fine-grained details. This dual approach ensures strong expressive power across various problem types.

A Chebyshev feature map is a specialized quantum feature map designed to encode classical data into quantum states using Chebyshev polynomials as the basis functions. The map is implemented using single-qubit rotation gates parameterized by a nonlinear function of the input variable $x\in\mathbb{R}$. The core building block is defined as
\begin{equation}\label{cheby}
\hat{\mathcal{U}}_\varphi(x) = 
\overset{N}{\underset{j=1}{\bigotimes}} 
\hat{R}_{Y,j}(\varphi(x)),
\end{equation}
where ${\hat{R}}_{Y,j}$ is a Pauli-$Y$ rotation gate on qubit $j$ ($j={1,\ldots,N}$), where $N$ is the number of qubits in the register,
\begin{equation}\label{rotation}
{\hat{R}}_{Y,j}=exp{\left(-i\frac{\varphi\left(x\right)}{2}Y_j\right)},
\end{equation}
and $\varphi(x)$ is a nonlinear encoding angle function.

\subsubsection{Chebyshev feature maps}\label{subsec2}

The encoding angle for the Chebyshev feature map is given by $\varphi(x)=2n[j] \arccos(x)$, where the function $n[j]$ assigns a polynomial degree to the $j$-th qubit. If $n[j]=1$ for all qubits, it is called a Sparse Chebyshev Feature Map. If $n[j]=j$ across qubits, it is a Chebyshev Tower Feature Map~\cite{kyriienko2021solving}.

When expanded using Euler's formula, the rotation gate ${\hat{R}}_{Y,j}(x)$ decomposes into
\begin{equation}
{\hat{R}}_{Y,j}(x)=T_n(x)I_j+\sqrt{1-x^2}U_{n-1}(x)X_jZ_j,
\end{equation}
where $T_n(x)$ is the Chebyshev polynomial of the first kind, and $U_n(x)$ is the Chebyshev polynomial of the second kind~\cite{trefethen2019approximation}. $X_j$, $Z_j$ are Pauli matrices.

This decomposition directly encodes Chebyshev polynomials into the quantum circuit, forming a comprehensive basis set for function approximation.

\subsubsection{Fourier feature maps}\label{subsubsec2}

The Fourier feature mapping is implemented on $N$ qubits, where each qubit serves as a basis for a specific frequency component. For an input variable $x\in\mathbb{R}$, the quantum state is prepared by applying a Pauli-$Y$ rotation gate ${\hat{R}}_{Y,j}$ to each qubit $j$ ($j={1,\ldots,N}$). The rotation angle for the $j$-th qubit is defined by the encoding angle function $\varphi_j\left(x\right)=2\pi\omega_jx$ in Eq.~(\ref{cheby}), where $\omega_j$ is the fundamental frequency assigned to the $j$-th qubit. Thus the overall Fourier feature map ${\hat{\mathcal{U}}}_\varphi\left(x\right)$ is
\begin{equation}\label{fourier}
\hat{\mathcal{U}}_\varphi(x) = \mathop{\bigotimes}_{j=1}^{N} \hat{R}_{Y,j}(2\pi\omega_j x).
\end{equation} 

By providing a diverse spectral representation of the input at the first layer of the quantum circuit, this encoding strategy facilitates the subsequent variational circuit in modeling complex functional dependencies governed by nonlinear differential equations, overcoming the limitations of simple angle encoding.

\subsection{Ansatz}\label{sec_ansatz}

The variational ansatz employed in this work is designed to balance expressivity and trainability for near-term quantum devices. The circuit consists of $L$ identical layers, each comprising parameterized single-qubit rotations followed by an entangling block with an alternating connectivity pattern.

Let $j \in \{1,\ldots,N\}$ index the $N$ qubits. The total number of tunable parameters is $3NL$, which are organized into a parameter matrix $\boldsymbol{\theta} \in \mathbb{R}^{L \times 3N}$. Each layer $\ell \; (\ell = 1,\ldots,L)$ is constructed as follows.

\textbf{Single-Qubit Rotations:} For each qubit $j$, a sequence of Euler rotations is applied,
\begin{equation}
    R_Z(\theta_{\ell,3j-2}) \; R_Y(\theta_{\ell,3j-1}) \; R_X(\theta_{\ell,3j}),
\end{equation}
where $\theta_{\ell, k}$ denotes the $k$-th parameter in the $\ell$-th row of $\boldsymbol{\theta}$. This sequence implements a general single-qubit rotation, ensuring rich local expressivity.

\textbf{Alternating Entanglement Pattern:} To create correlations between qubits while maintaining hardware-friendly connectivity, the entangling gates alternate between two patterns across layers:
\begin{itemize}
    \item \textbf{Odd layers ($\ell$ odd)}: A linear chain topology, applying CNOT gates between nearest neighbors:
    \begin{equation}
        \mathrm{CNOT}(j, j+1) \quad \text{for } j = 1, \ldots, N-1.
    \end{equation}
    \item \textbf{Even layers ($\ell$ even)}: A ring (cyclic) topology, which also connects the last qubit back to the first:
    \begin{equation}
        \mathrm{CNOT}(j, j+1 \text{ mod } N) \quad \text{for } j = 1, \ldots, N,
    \end{equation}
    where qubit index $N+1$ is identified with qubit $1$.
\end{itemize}

\begin{figure}[htb]
\centering
\includegraphics[width=\linewidth]{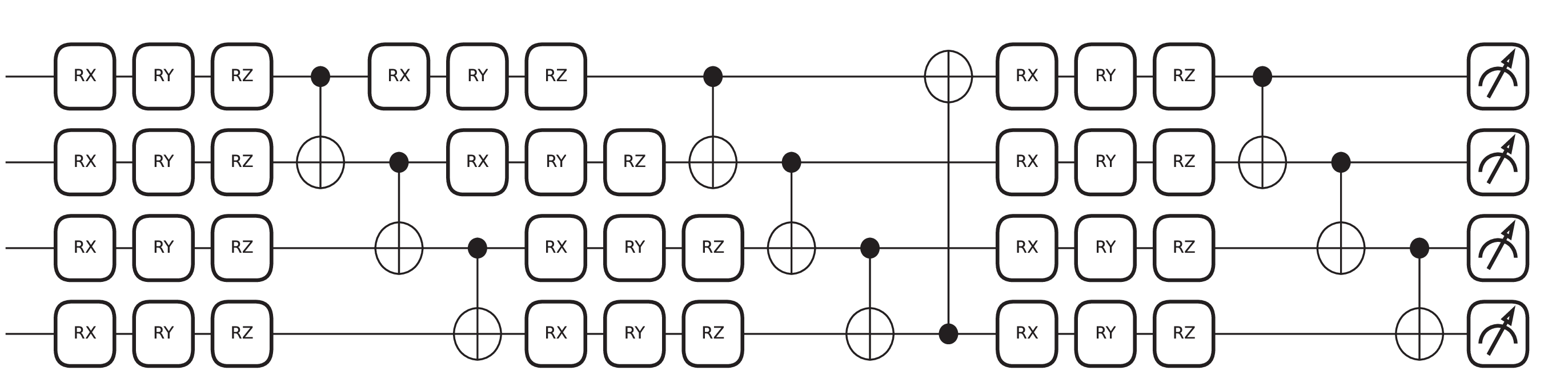}
\caption{Quantum Ansatz with alternating entanglement pattern for 4 qubits and 3 layers.}\label{ansatz}
\end{figure}

The ansatz is illustrated in Fig.~\ref{ansatz}. The ansatz alternates between linear and ring connectivity, which enhances the entanglement capability of the circuit and facilitates the flow of information across all qubits, which is crucial for modeling complex correlations in the solutions of differential equations. The ansatz is compatible with the limited connectivity of many NISQ devices while maintaining the potential for generating highly entangled states necessary for expressive function approximation.

\subsection{Hard Constraint Embedding}

We consider a first-order ordinary differential equation (ODE) with an initial condition,
\begin{equation}\label{ode1}
\frac{{dy(x)}}{{dx}} = g(x,y(x)), \quad x \in [a,b],
\end{equation}
and the initial condition $y(a) = A$.

To solve the ODE, we construct a trial solution $\widetilde{y}(x)$ that strictly and automatically satisfies the initial condition. This is the core idea of the hard constrained embedding method. The trial solution takes the form,
\begin{equation}
 \ \widetilde{y}(x) = A + K(x)f(\boldsymbol{\theta},x),
\end{equation}
where $K\left(x\right)$ is a kernel function chosen to enforce the boundary condition, and $f(\boldsymbol{\theta},x)$ is the raw output of a parameterized quantum model with input $x$ and ansatz parameters $\boldsymbol{\theta}$. By design, $K\left(a\right)=0$, which guarantees $\widetilde{y}(a) =A$ regardless of $f(\boldsymbol{\theta},a)$, thereby embedding the initial condition as a hard constraint.

\subsubsection{Linear hard constrained embedding}

A straightforward and universally applicable choice is the linear kernel $K\left(x\right)=x-a$, leading to the trial solution,
\begin{equation}\label{linearker}
 \widetilde{y}(x) = A + (x - a)f(\boldsymbol{\theta} ,x).
\end{equation}

This linear embedding is the foundational approach. It ensures the initial condition is satisfied by construction while allowing the quantum model $f(\boldsymbol{\theta},x)$ to learn the derivative field freely. Its simplicity offers excellent trainability and serves as a robust baseline for a wide range of problems.

\subsubsection{Exponential hard constrained embedding}

While the linear hard constraint embedding provides a solid foundation, it has limitations for many applications where systems exhibit saturating nonlinear behavior (e.g., population growth, drug concentration, capacitor charging). Therefore, we also construct an exponential hard constrained embedding by the exponential kernel $(1 - {e^{ - (x - a)}})$,
\begin{equation}
 \widetilde{y}(x) = A + (1 - {e^{ - (x - a)}})f(\boldsymbol{\theta},x).
\end{equation}

The exponential function inherently captures a rapid initial change that gradually saturates---a ubiquitous pattern in nature that the linear embedding cannot naturally represent. Choosing the exponential embedding is a strategic method of incorporating domain-specific knowledge. It exemplifies a key principle in physics-informed machine learning: embedding domain knowledge into the model architecture through well-chosen basis functions reduces the hypothesis space, leading to more efficient and physically plausible solutions.

\subsection{Derivative computation via parameter-shift rule}

The derivatives of the trial solution with respect to the input variable $x$ are computed using the parameter-shift rule~\cite{schuld2019evaluating}, which enables exact gradient calculation for quantum circuits with generator-based parameterizations.

\subsubsection{Mathematical Formulation}

For a trial solution defined as $\widetilde{y}(x) = A + K(x)f(\boldsymbol{\theta},x)$, where $K\left(x\right)$ is a kernel function and $f(\boldsymbol{\theta},x)$ represents the quantum model output, its derivative with respect to $x$ is given by
\begin{equation}\label{parashift1}
\frac{\partial \widetilde{y}(x)}{\partial x}=\frac{dK(x)}{dx}\cdot f(\boldsymbol{\theta},x)+K(x)\cdot\frac{\partial f(\boldsymbol{\theta},x)}{\partial x}.
\end{equation}
Here $K\left(x\right)$ takes the form $K\left(x\right)=x-a$ for linear embedding, or $K(x)=1-e^{-(x-a)}$ for exponential embedding.

\subsubsection{Quantum derivative computation for encoding function}

\begin{figure}[htb]
\centering
\includegraphics[width=\linewidth]{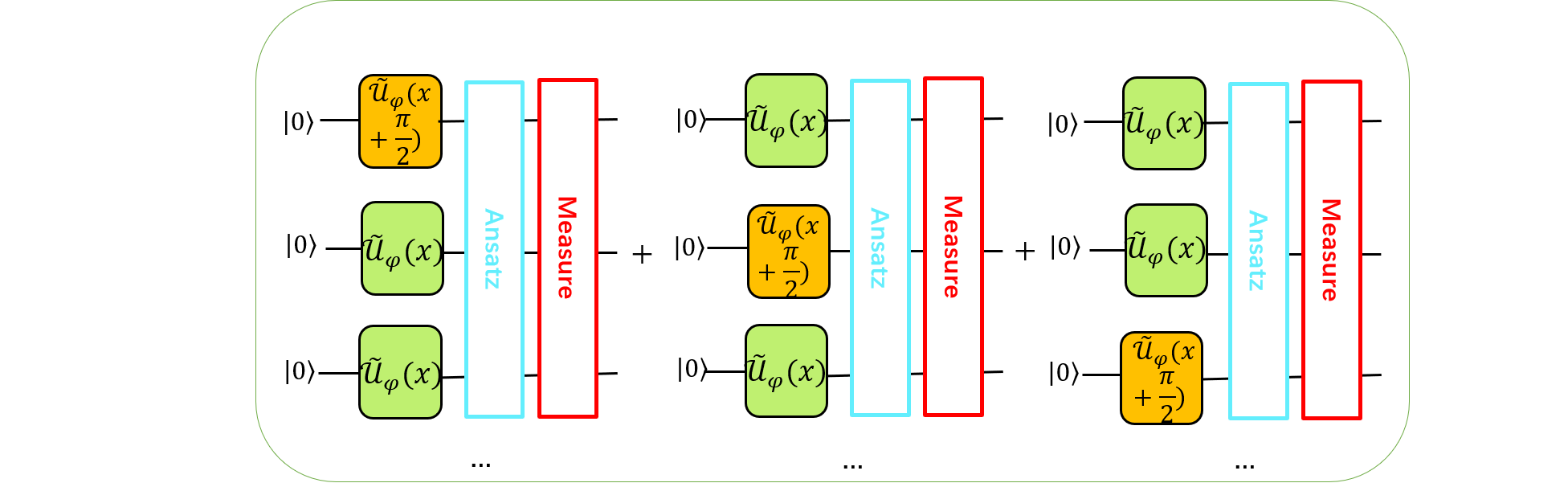}
\caption{Derivative computation via parameter-shift rule.}\label{parashift}
\end{figure}

The quantum gradient $\frac{\partial f(\boldsymbol{\theta},x)}{\partial x}$ is computed using the parameter-shift rule. When the input $x$ is encoded via a nonlinear encoding angle function $\varphi(x)$, the derivative is computed as
\begin{equation}\label{parashift2}
\frac{\partial f(\boldsymbol{\theta},x)}{\partial x}=\frac{1}{2}[f({\hat{\mathcal{U}}}_\varphi\left(\varphi(x)+\frac{\pi}{2}\right))-f({\hat{\mathcal{U}}}_\varphi\left(\varphi(x)-\frac{\pi}{2}\right))],
\end{equation}
where $\varphi\left(x\right)$ is the encoding angle function. For Chebyshev feature maps, $\varphi(x)=2n[j] \arccos(x)$, while for Fourier feature maps, $\varphi(x)=2\pi\omega_jx$. The derivative computation via parameter-shift rule is illustrated in Fig.~\ref{parashift}.

\subsubsection{Computational cost analysis}

We note that the parameter-shift rule, while providing exact gradients, incurs a computational cost that scales with the number of circuit evaluations. Specifically, computing the derivative $\partial f/\partial x$ at each collocation point requires two additional circuit evaluations (corresponding to $\varphi(x) \pm \pi/2$), as shown in Eq.~(\ref{parashift2}). For $M$ collocation points, the derivative computation thus requires $2M$ extra circuit evaluations per training step. Furthermore, the gradients of the loss with respect to each trainable parameter $\theta_k$ also require two circuit evaluations per parameter per collocation point. For a circuit with $P$ trainable parameters, the total number of circuit evaluations per training step scales as $\mathcal{O}(M \cdot (1 + 2 + 2P))$, where the terms correspond to the function evaluation, the derivative via parameter-shift, and the parameter gradients, respectively. In contrast, classical automatic differentiation typically incurs a constant-factor overhead relative to the forward pass. We acknowledge that this cost analysis does not establish a computational advantage of the quantum approach over classical methods; rather, it is provided to clarify the resource requirements of the proposed framework. A rigorous comparison of wall-clock runtime and sample complexity between quantum and classical approaches on equivalent problems remains an important topic for future investigation.

\subsection{Loss function}

For solving differential equations using quantum models, we employ a physics-informed loss function that systematically incorporates the differential equation residual and reference data fitting. The loss function is designed with adaptive weighting to balance different constraints during training.

The comprehensive loss function is defined as
\begin{equation}
\mathcal{L}_{\mathrm{total}}(\mathbf{\Theta}) = 
\frac{1}{M} \sum_{i=1}^{M} \left[ \lambda_{\mathrm{res}} L_{\mathrm{res}}^{(i)}(\mathbf{\Theta}) + \lambda_{\mathrm{data}} L_{\mathrm{data}}^{(i)}(\mathbf{\Theta}) \right],
\end{equation}
where $\mathbf{\Theta}$ represents the complete set of trainable parameters (including quantum circuit parameters $\boldsymbol{\theta}$, output weights $\overrightarrow t$, and bias term $\overrightarrow b$). $M$ is the number of training collocation points. $\lambda_{\mathrm{res}}$ and $\lambda_{\mathrm{data}}$ are adaptive weights. These coefficients can be dynamically adjusted during training based on the magnitude of each loss component, preventing any single term from dominating the optimization process.

\textbf{Differential Equation Residual Loss}
\begin{equation}
\mathcal{L}_{\mathrm{res}}^{\left(i\right)}\left(\mathbf{\Theta}\right)=\left[\frac{d \widetilde{y}(x)}{dx}-g\left(x, \widetilde{y}(x)\right)\right]^2,
\end{equation}
where $\widetilde{y}(x) = A + K(x) f(\boldsymbol{\theta}, x)$ is the trial solution that strictly enforces the initial/boundary conditions. The derivative $\frac{d \widetilde{y}(x)}{dx}$ is computed using the parameter-shift rule as described in Eq.~(\ref{parashift1}).

\textbf{Reference data Fitting Loss}
\begin{equation}
\mathcal{L}_{\mathrm{data}}^{(i)}(\mathbf{\Theta})=[ \widetilde{y}(x)-y_{\mathrm{cla}}(x)]^2,
\end{equation}
where ${y_{\mathrm{cla}}}(x)$ denotes the classical numerical or analytical reference solution used to guide the quantum model toward physically meaningful solutions.

The complete workflow of our physics-informed quantum machine learning approach with hard constraint embedding is summarized in Algorithm~1.

\begin{algorithm}[Physics-Informed Quantum Machine Learning with Hard Constraint Embedding]
\label{algo:quantum_ode_solver}
\begin{algorithmic}[1]
\Require 
\Statex Differential equation: $\frac{dy}{dx} = g(x, y)$
\Statex Initial condition: $y(a) = A$
\Statex Domain: $[a, b]$
\Statex Number of qubits: $N$
\Statex Number of ansatz layers: $L$
\Statex Feature map type: $\mathcal{F} \in \{\text{Chebyshev}, \text{Fourier}\}$
\Statex Embedding kernel: $\mathcal{K} \in \{\text{Linear}, \text{Exponential}\}$
\Statex Reference data (optional): $\{(x_i, y_{\mathrm{ref}}(x_i))\}_{i=1}^{M_{\mathrm{data}}}$
\Ensure 
\Statex Optimized parameters $\boldsymbol{\Theta}^*$
\Statex Trial solution $\widetilde{y}(x; \boldsymbol{\Theta}^*)$

\State \textbf{Initialization}
\State Generate collocation points $\{x_i\}_{i=1}^{M} \subset [a, b]$ 
\State Initialize all parameters $\boldsymbol{\Theta} = \{\boldsymbol{\theta}, \mathbf{w}, b\}$
\State Set adaptive weights $\lambda_{\mathrm{res}}, \lambda_{\mathrm{data}}$ 
\State Construct kernel function $K(x)$ based on $\mathcal{K}$ 
\State Define encoding function $\varphi(x)$ based on $\mathcal{F}$ 

\State \textbf{Quantum Circuit Evaluation Loop}
\For{each collocation point $x_i$}
    \State \textbf{State Preparation:}
    \State Apply feature map: $\lvert \psi_{\mathrm{in}}(x_i) \rangle = \hat{\mathcal{U}}_\varphi(x_i) \lvert 0 \rangle^{\otimes N}$
    \State \textbf{Quantum Variational Evolution:}
    \State Apply ansatz: $\lvert \psi_{\boldsymbol{\theta}}(x_i) \rangle = V(\boldsymbol{\theta}) \lvert \psi_{\mathrm{in}}(x_i) \rangle$
    \State \textbf{Observable Measurement:}
    \State Compute $f(\boldsymbol{\theta}, x_i) = \sum_{j=1}^{N} w_j \langle Z_j \rangle_{\boldsymbol{\theta}, x_i} + b$
    \State \textbf{Trial Solution \& Derivative:}
    \State $\widetilde{y}(x_i) = A + K(x_i) f(\boldsymbol{\theta}, x_i)$
    \State Compute $\partial f(\boldsymbol{\theta}, x_i)/\partial x$ via parameter-shift rule on $\varphi(x_i)$ using Eqs.~(\ref{parashift1})(\ref{parashift2})
    \State \textbf{Loss Accumulation:}
    \State $\mathcal{L}_{\mathrm{res}}^{(i)} = [\partial \widetilde{y}(x_i)/\partial x - g(x_i, \widetilde{y}(x_i))]^2$
    \State $\mathcal{L}_{\mathrm{data}}^{(i)} = [\widetilde{y}(x_i) - y_{\mathrm{cla}}(x_i)]^2$ 
\EndFor

\State \textbf{Loss Computation \& Optimization}
\State $\mathcal{L}_{\mathrm{total}}(\boldsymbol{\Theta}) = \frac{1}{M} \sum_{i=1}^{M} [\lambda_{\mathrm{res}} \mathcal{L}_{\mathrm{res}}^{(i)} + \lambda_{\mathrm{data}} \mathcal{L}_{\mathrm{data}}^{(i)}]$
\State Update $\boldsymbol{\Theta} \gets \boldsymbol{\Theta} - \eta \nabla_{\boldsymbol{\Theta}} \mathcal{L}_{\mathrm{total}}(\boldsymbol{\Theta})$ \Comment{Using gradient-based optimizer}
\State Optionally update $\lambda_{\mathrm{res}}, \lambda_{\mathrm{data}}$ based on loss magnitudes

\While{not converged \textbf{and} $t < t_{\max}$}
    \State Repeat steps 4--6
    \State Check convergence: $|\mathcal{L}_{\mathrm{total}}^{(t+1)} - \mathcal{L}_{\mathrm{total}}^{(t)}| < \epsilon$
\EndWhile

\State \Return $\boldsymbol{\Theta}^*, \widetilde{y}(x; \boldsymbol{\Theta}^*)$
\end{algorithmic}
\end{algorithm}

\begin{figure}[htb]
\centering
\includegraphics[width=0.8\linewidth]{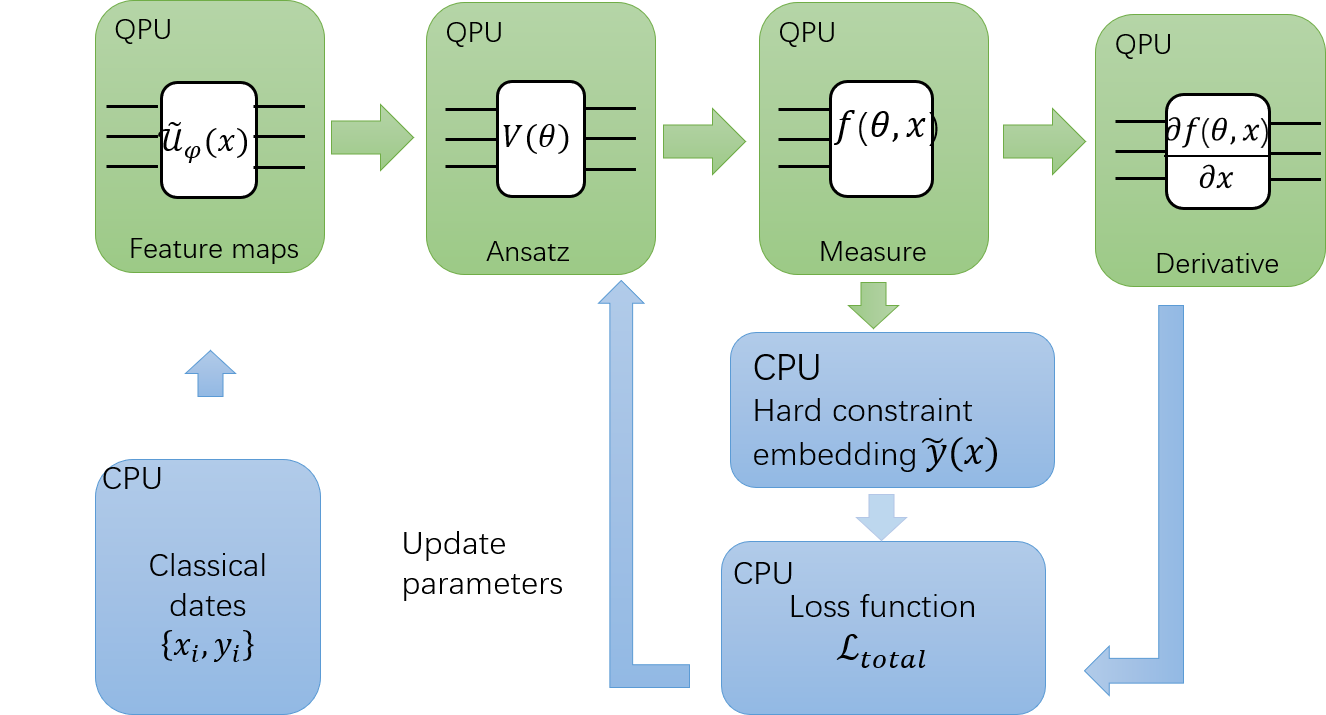}
\caption{Flowchart of the proposed PIQML framework with hard constraint embedding.}\label{quanliucheng}
\end{figure}

\section{Results}\label{sec3}

\subsection{First-order ODE}

We first considered a simple linear ODE with a known analytical solution to facilitate rigorous error analysis,
\begin{equation}\label{efirstorder}
\frac{{dy}}{{dx}} = 2x - \frac{{x{e^{\frac{{ - {x^2}}}{2}}}}}{{{x^3} + x + 1}} - \frac{{(3{x^2} + 1){e^{\frac{{ - {x^2}}}{2}}}}}{{{{({x^3} + x + 1)}^2}}},\quad y(0) = 1.
\end{equation}
The analytic solution is $y = {x^2} + \frac{{{e^{\frac{{ - {x^2}}}{2}}}}}{{{x^3} + x + 1}}$.

\begin{figure}[htb]
\centering
\subfloat[]{%
\includegraphics[width=0.48\linewidth]{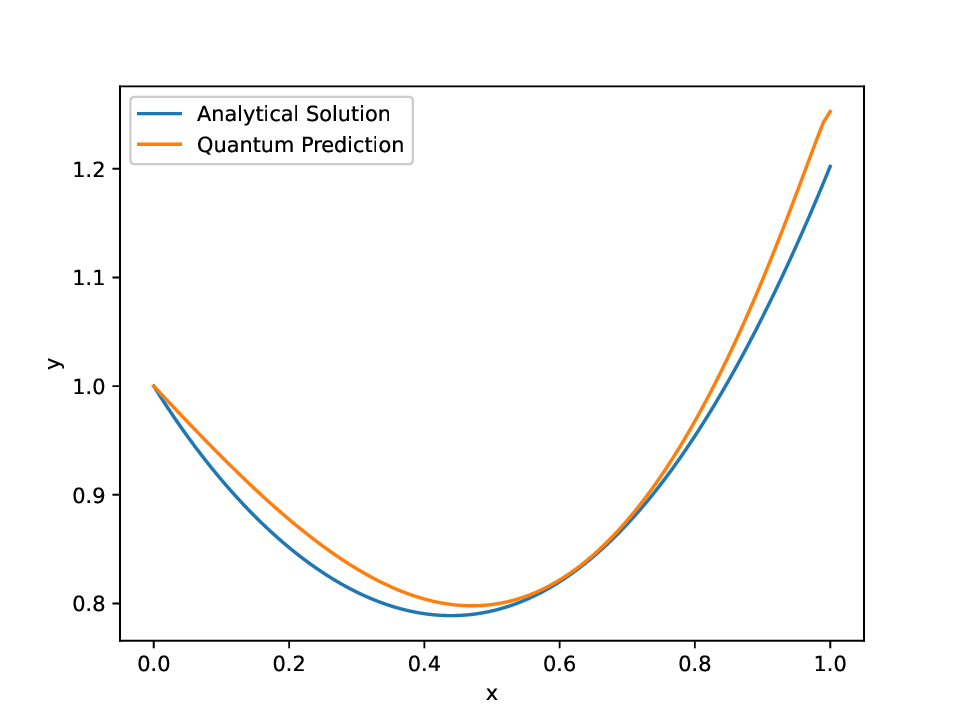}%
}%
\hfill%
\subfloat[]{%
\includegraphics[width=0.48\linewidth]{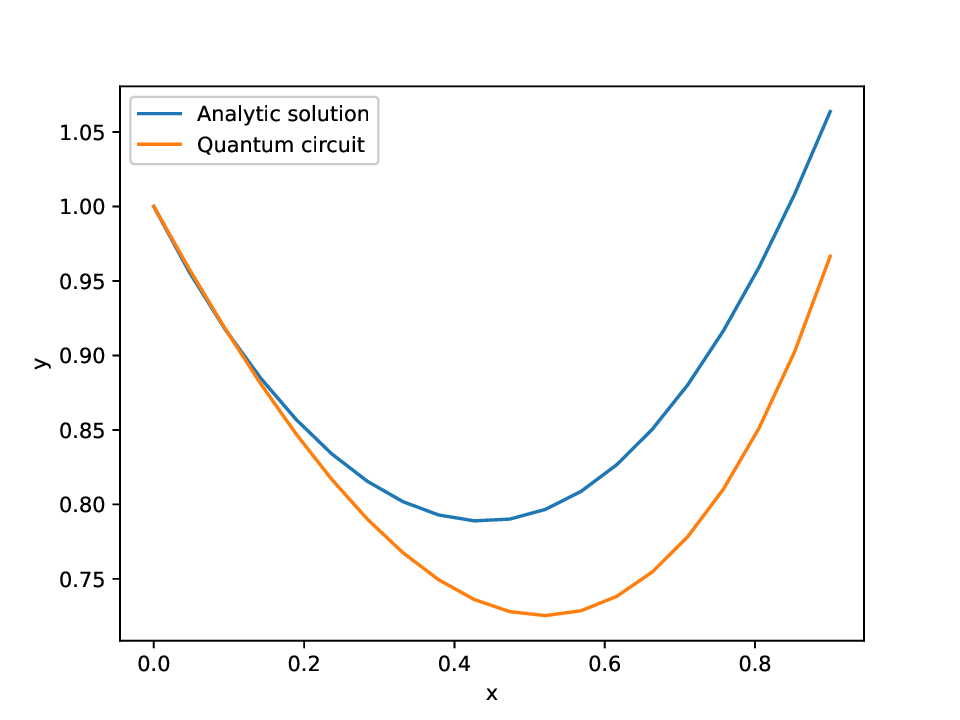}%
}
\caption{(\textbf{a}) Comparison of the PIQML model (orange line) against the analytical solution (blue line) for a simple linear ODE. (\textbf{b}) Performance of an existing quantum method from the literature (orange line)~\cite{kyriienko2021solving} in fitting the same ODE, showing inferior accuracy compared to the PIQML model.}
\label{firstorder}
\end{figure}

We employ a linear hard constraint to enforce the initial condition $\widetilde{y}(x) = 1 + (x - 0)f(\boldsymbol{\theta} ,x)$. The results are presented in Fig.~\ref{firstorder} and Table~\ref{tfirstorder}. Fig.~\ref{firstorder}(\textbf{a}) reveals that our PIQML model (orange line) aligns closely with the analytical solution (blue line). Conversely, Fig.~\ref{firstorder}(\textbf{b}) illustrates the performance of an existing quantum method from the literature (orange line)~\cite{kyriienko2021solving}, which employs a differentiable quantum circuit approach without hard constraint embedding, using automatic differentiation for gradient computation instead of the parameter-shift rule, which shows inferior fitting performance compared to our PIQML model.

\begin{figure}[htb]
\centering
\subfloat[]{%
\includegraphics[width=0.48\linewidth]{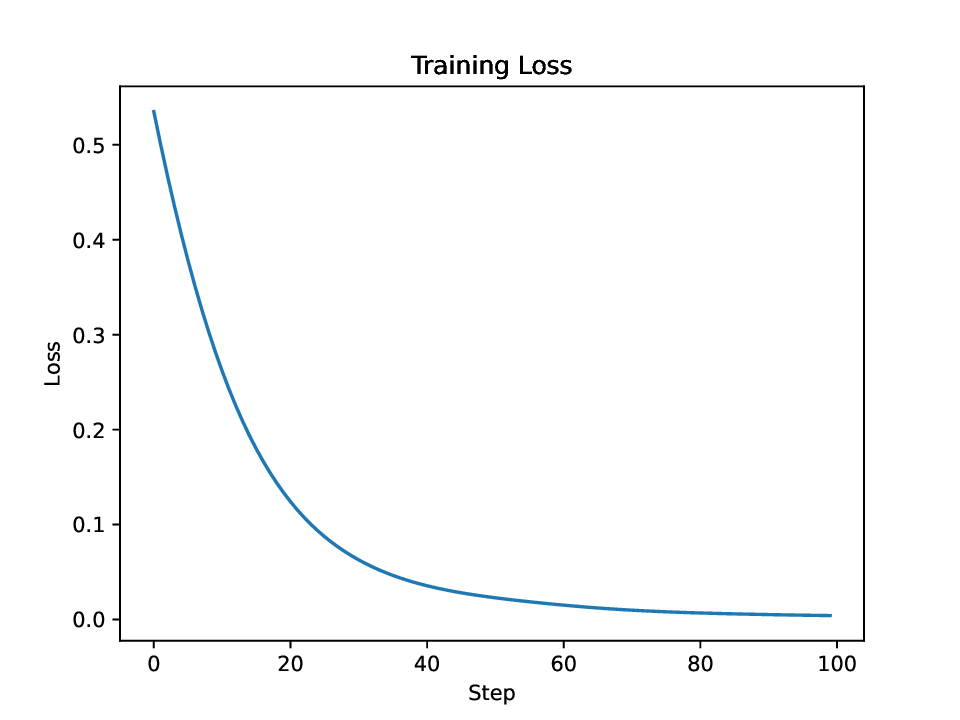}%
}%
\hfill%
\subfloat[]{%
\includegraphics[width=0.48\linewidth]{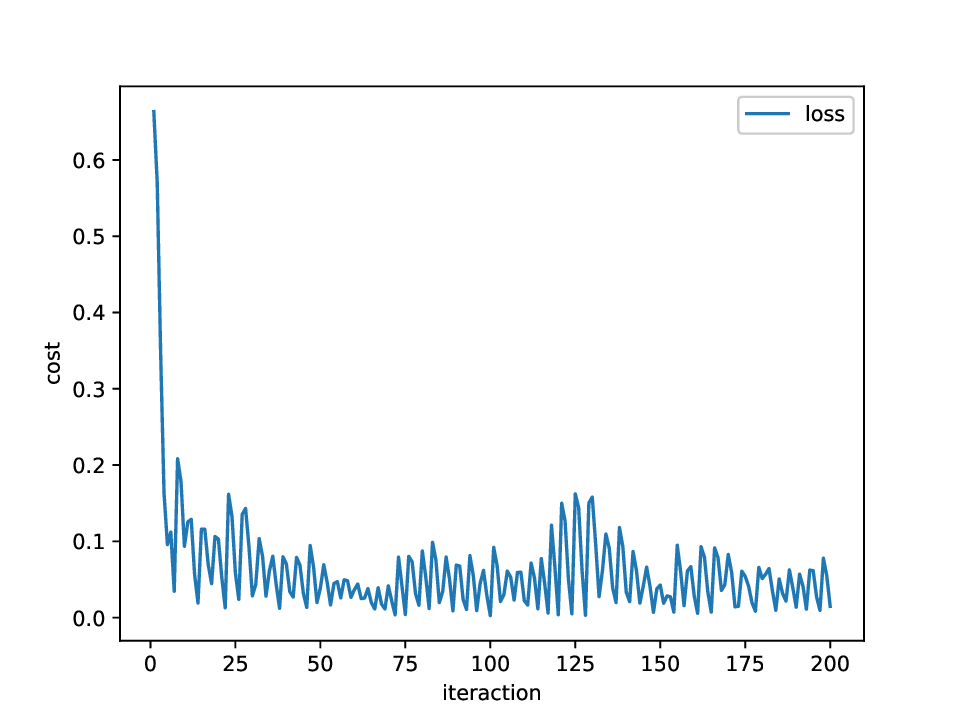}%
}
\caption{Comparison of the loss functions for a simple linear ODE. (\textbf{a}) Our PIQML model. (\textbf{b}) The existing quantum method from the literature~\cite{kyriienko2021solving}.}
\label{firstorderloss}
\end{figure}

The required number of qubits, quantum layers, and other parameters for the experiment are detailed in Table~\ref{tfirstorder}. The comparison highlights the efficiency of our approach. Using only 2 qubits and a 2-layer circuit (8 parameters total), our PIQML model converged within 100 optimization steps. In contrast, the reference method, despite having slightly fewer parameters (7 parameters total), required 200 steps to reach comparable accuracy. This accelerated convergence stems from our hard constraint embedding. Moreover, the adoption of the parameter-shift rule for quantum-native gradient computation circumvents the potential numerical instability and extra computational cost associated with automatic differentiation. The training loss evolution for both methods is also reported in Fig.~\ref{firstorderloss}, confirming that our PIQML model achieves a lower final loss.

\begin{table*}[htbp]
\centering
\caption{Comparison of model configurations for Eq.~(\ref{efirstorder}).}\label{tfirstorder}
\begin{tabular}{@{}lcc@{}}
\toprule
& \textbf{PIQML (Ours)} & \textbf{Ref.~\cite{kyriienko2021solving}} \\
\midrule
Number of qubits     & 2 & 2 \\
Number of layers     & 2 & 1 \\
Number of parameters & 8 & 7 \\
Training steps       & 100 & 200 \\
Loss function        & Physics-informed & Physics-informed \\
Hard constraint      & Yes & No \\
Gradient computation & Parameter-shift & Automatic differentiation \\
\bottomrule
\end{tabular}
\end{table*}

To further demonstrate the superiority of our PIQML framework, we conducted a comprehensive baseline comparison against several classical physics-informed machine learning methods for solving Eq.~(\ref{efirstorder}). The classical baselines include: (i)~a standard PINN with two hidden layers of 64 neurons; (ii)~PINN with hard constraint (PINN\_HC), which embeds the initial condition analytically; (iii)~Adaptive PINN with a learnable loss weight; (iv)~DeepONet-style architecture; (v)~Fourier Neural Operator (FNO)-style architecture; and (vi)~Spectral PINN with 16 Fourier coefficients. All classical models were trained for 200 epochs using the Adam optimizer with a learning rate of $10^{-3}$, while our PIQML model employs 2 qubits, 2 variational layers (11 total parameters), and 100 training steps.

\begin{figure*}[htb]
\centering
\includegraphics[width=\linewidth]{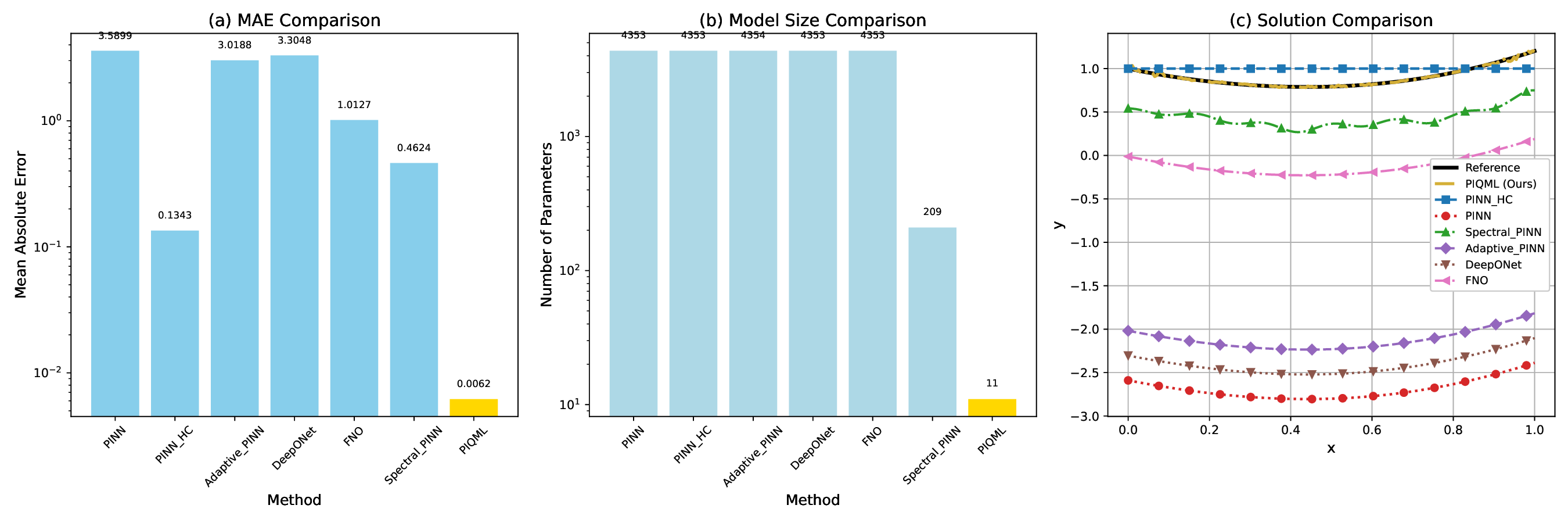}
\caption{Baseline comparison between classical physics-informed machine learning methods and the proposed PIQML model for Eq.~(\ref{efirstorder}). (\textbf{a}) Mean Absolute Error (MAE) for each method on a logarithmic scale; the PIQML model achieves the lowest error. (\textbf{b}) Number of trainable parameters for each model on a logarithmic scale; PIQML requires significantly fewer parameters than all classical baselines. (\textbf{c}) Solution comparison: analytical reference (solid black line) versus predictions from all methods, with the PIQML model showing the closest agreement to the analytical solution.}\label{baselinecomp}
\end{figure*}

The results are summarized in Fig.~\ref{baselinecomp}. Fig.~\ref{baselinecomp}(\textbf{a}) compares the Mean Absolute Error (MAE) across all methods on a logarithmic scale. The PIQML model achieves the lowest MAE, outperforming all classical baselines despite using far fewer parameters. Fig.~\ref{baselinecomp}(\textbf{b}) presents the model size comparison in terms of trainable parameters (logarithmic scale). The PIQML model requires only 11 parameters---over two orders of magnitude fewer than the classical methods (e.g., PINN: 4{,}353 parameters, DeepONet: 4{,}353 parameters, Spectral PINN: 209 parameters). Fig.~\ref{baselinecomp}(\textbf{c}) shows the solution profiles for all methods alongside the analytical reference. The PIQML prediction closely follows the analytical solution, while classical methods exhibit varying degrees of deviation, particularly in regions with higher curvature.

\subsection{First-order Polynomial ODE}

\begin{figure}[htb]
\centering
\subfloat[]{%
\includegraphics[width=0.48\linewidth]{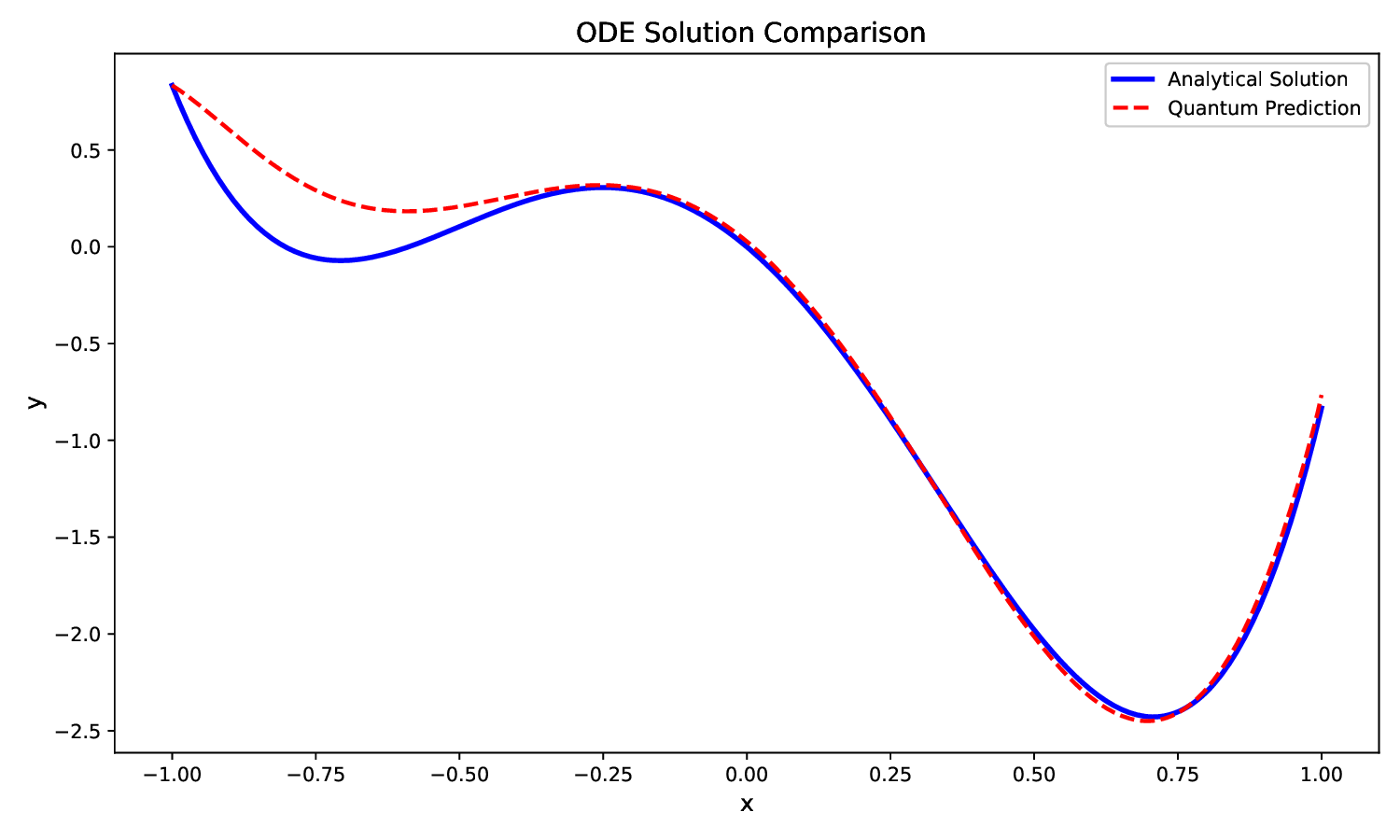}%
}%
\hfill%
\subfloat[]{%
\includegraphics[width=0.48\linewidth]{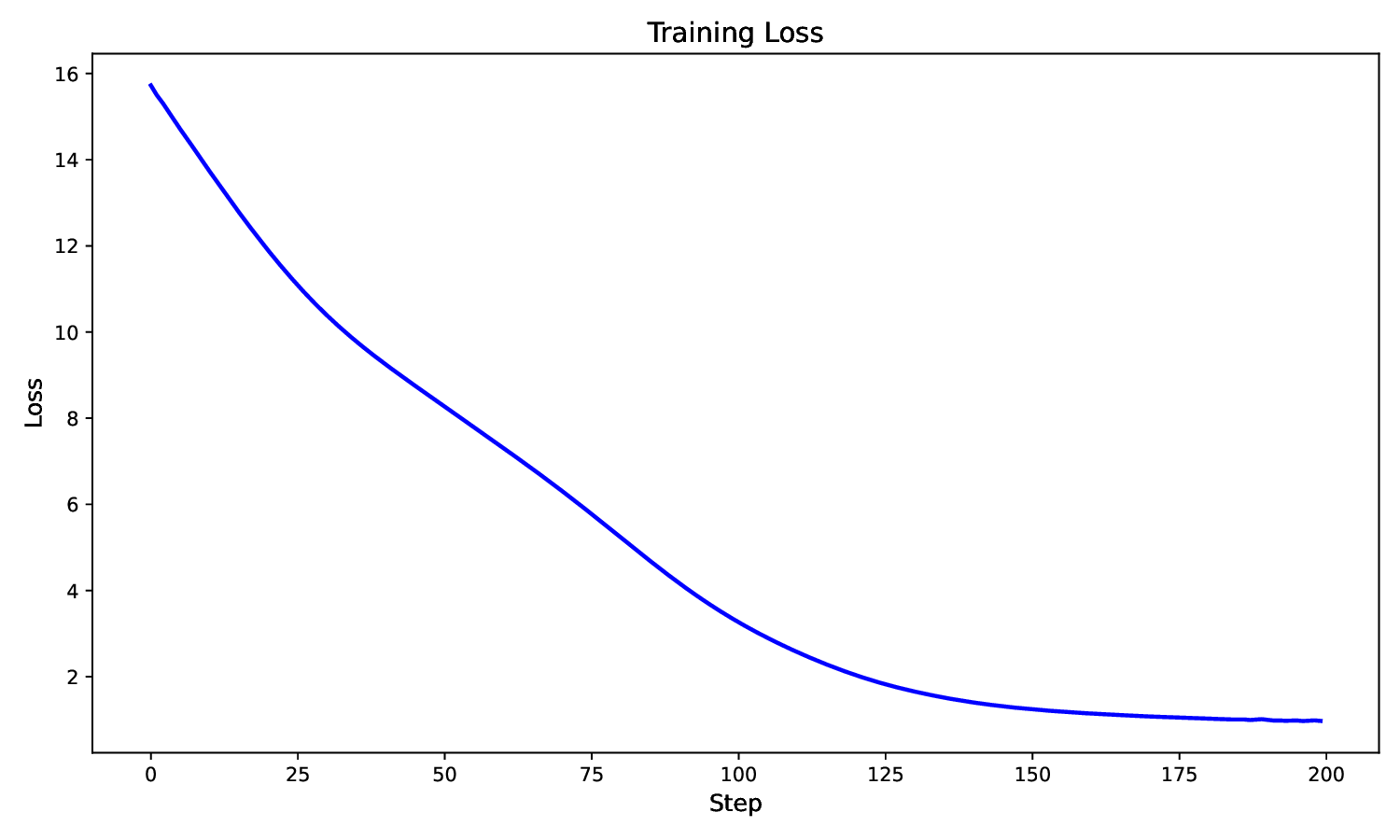}%
}

\vspace{6pt}

\subfloat[]{%
\includegraphics[width=0.48\linewidth]{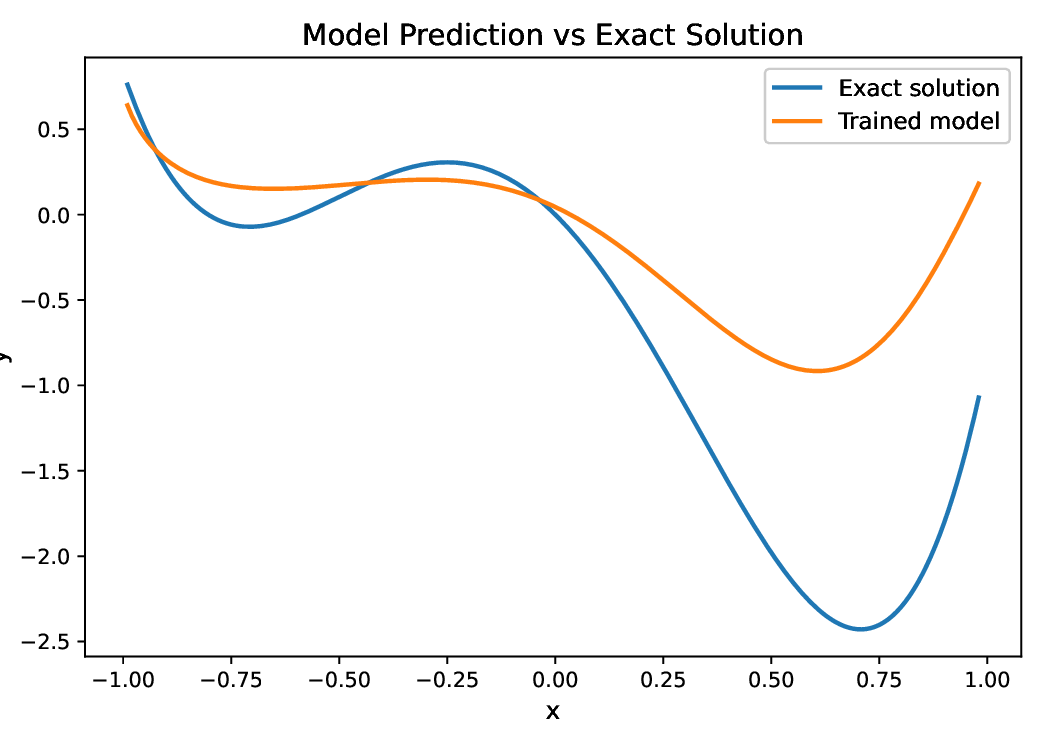}%
}%
\hfill%
\subfloat[]{%
\includegraphics[width=0.48\linewidth]{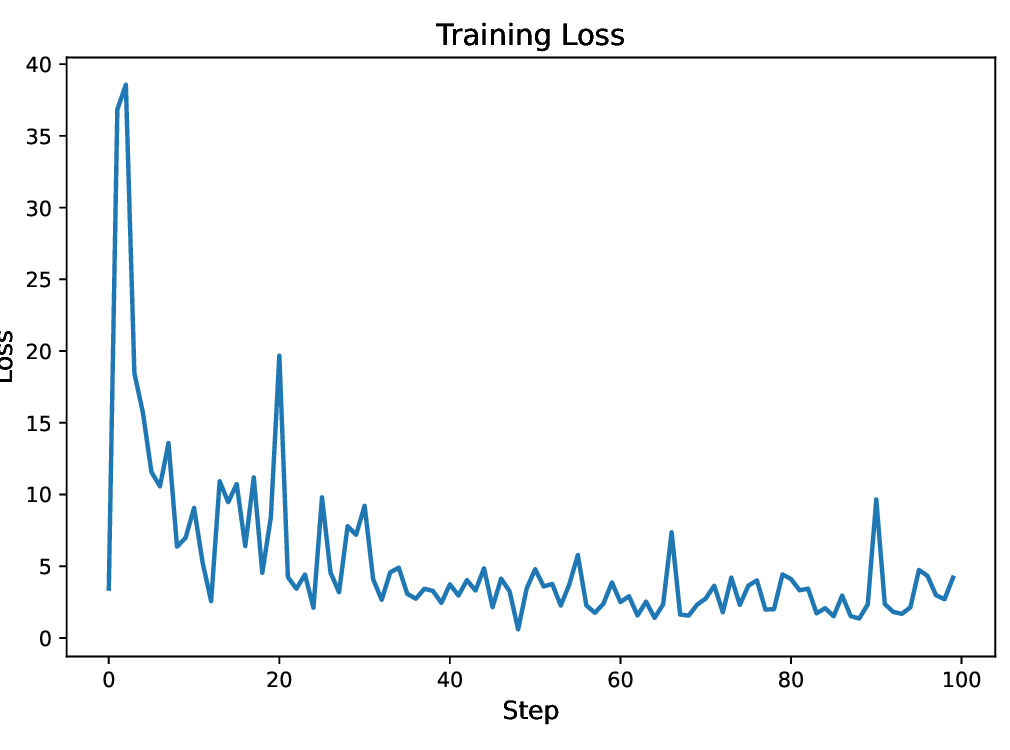}%
}
\caption{(\textbf{a}) Comparison of the solution to a first-order polynomial ODE (blue line) and the PIQML model prediction (orange line), demonstrating the model's ability to capture nonlinear oscillatory dynamics. (\textbf{b}) Convergence of the physics-informed loss function for the PIQML model, indicating stable and effective training. (\textbf{c}) Comparison of the solution to the same ODE (blue line) and an existing quantum method from the literature (orange line)~\cite{mitarai2018quantum}. (\textbf{d}) The training loss for the existing quantum method~\cite{mitarai2018quantum}.}
\label{fistpoly}
\end{figure}

For a first-order polynomial ODE,
\begin{equation}\label{efistpoly}
\frac{{dy}}{{dx}} = 5(4{x^3} + {x^2} - 2x - \frac{1}{2}),\quad y( - 1) = 5/6,
\end{equation}
the analytic solution is $y = 5({x^4} + \frac{1}{3}{x^3} - {x^2} - \frac{1}{2}x)$. The solution to this equation exhibits complex nonlinear oscillatory behavior as shown by the blue line in Fig.~\ref{fistpoly}(\textbf{a}). Instead of opting for the linear hard constrained embedding Eq.~(\ref{linearker}), we constructed an exponential hard constraint embedding
\begin{equation}
y(x) = \frac{5}{6} + (1 - {e^{ - (x + 1)}})f(\boldsymbol{\theta} ,x) , \quad -1\le x\le1,
\end{equation}
to strictly enforce the initial condition while allowing flexible fitting through the trainable function $f(\boldsymbol{\theta}, x)$.

As shown in Fig.~\ref{fistpoly}(\textbf{a}), our model successfully captured this dynamic, with the prediction curve (orange line) closely following the reference solution (blue line). The physics-informed loss function is presented in Fig.~\ref{fistpoly}(\textbf{b}), demonstrating effective convergence. The final loss is 0.9. For comparison, Fig.~\ref{fistpoly}(\textbf{c}) presents an existing quantum method without physics-informed regularization from the literature~\cite{mitarai2018quantum}. The predicted solution (orange curve in Fig.~\ref{fistpoly}(\textbf{c})) deviates noticeably from the reference, and the loss curve in Fig.~\ref{fistpoly}(\textbf{d}) remains higher and less stable. A direct comparison between Fig.~\ref{fistpoly}(\textbf{a}),(\textbf{c}) and Fig.~\ref{fistpoly}(\textbf{b}),(\textbf{d}) confirms that our Physics-Informed Quantum Machine Learning (PIQML) model achieves better accuracy and a significantly lower final loss.

The comparison in Table~\ref{tfistpoly} further highlights the efficiency of our approach. While the reference method required 3 qubits, a 3-layer circuit (27 parameters), and 100 optimization steps to solve this nonlinear problem, our PIQML model achieved higher accuracy using only 2 qubits, a 2-layer circuit (8 parameters), and 100 steps. This result strongly demonstrates the resource efficiency and enhanced expressivity of our PIQML framework. The hard constraint embedding avoids the repeated penalty for the initial condition during optimization.

\begin{table*}[htbp]
\centering
\caption{Comparison of model configurations for Eq.~(\ref{efistpoly}).}\label{tfistpoly}
\begin{tabular}{@{}lcc@{}}
\toprule
& \textbf{PIQML (Ours)} & \textbf{Ref.~\cite{mitarai2018quantum}} \\
\midrule
Number of qubits     & 2 & 3 \\
Number of layers     & 2 & 3 \\
Number of parameters & 8 & 27 \\
Training steps       & 100 & 100 \\
Loss function        & Physics-informed & None \\
Hard constraint      & Yes & No \\
Gradient computation & Parameter-shift & Automatic differentiation \\
\bottomrule
\end{tabular}
\end{table*}

\begin{figure}[htb]
\centering
\subfloat[]{%
\includegraphics[width=0.48\linewidth]{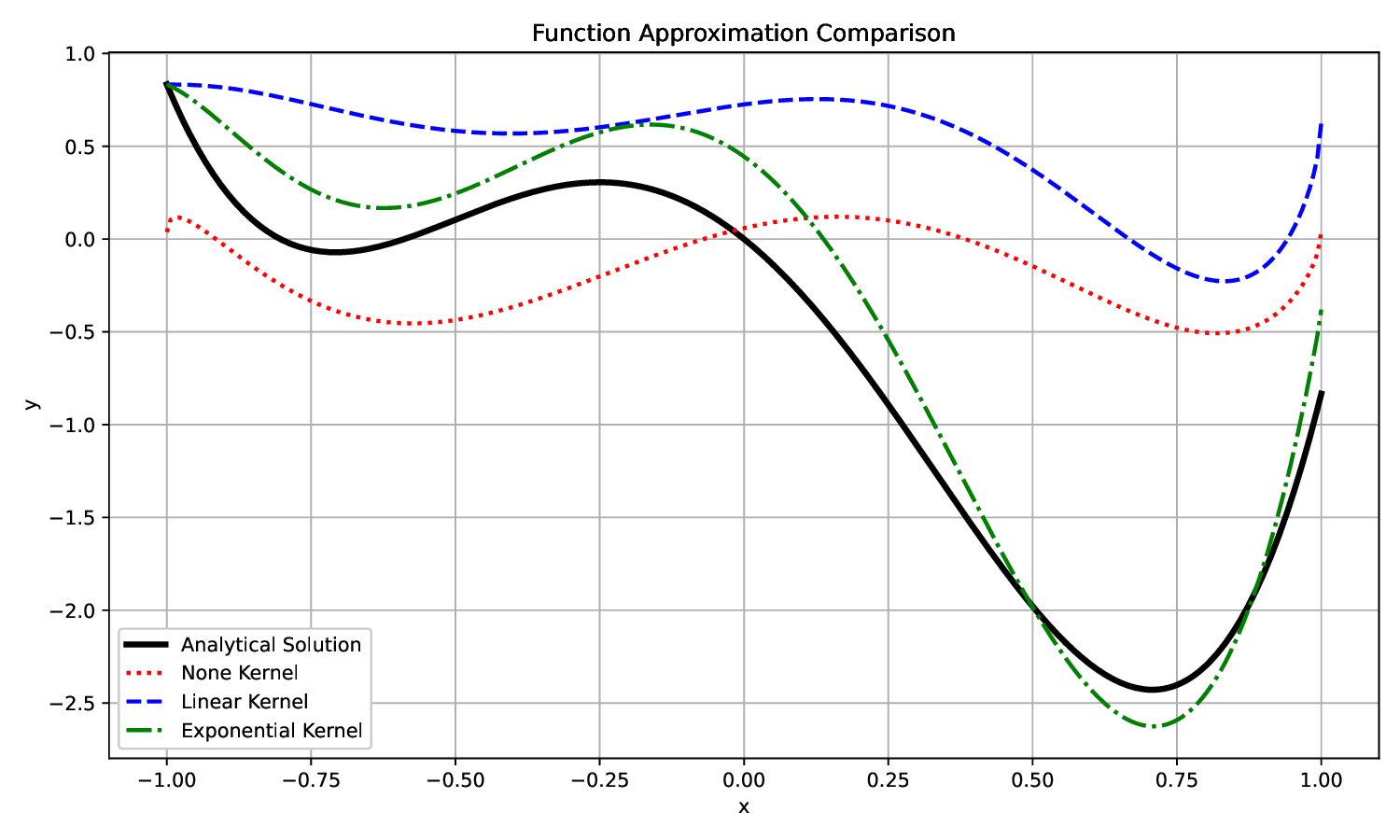}%
}%
\hfill%
\subfloat[]{%
\includegraphics[width=0.48\linewidth]{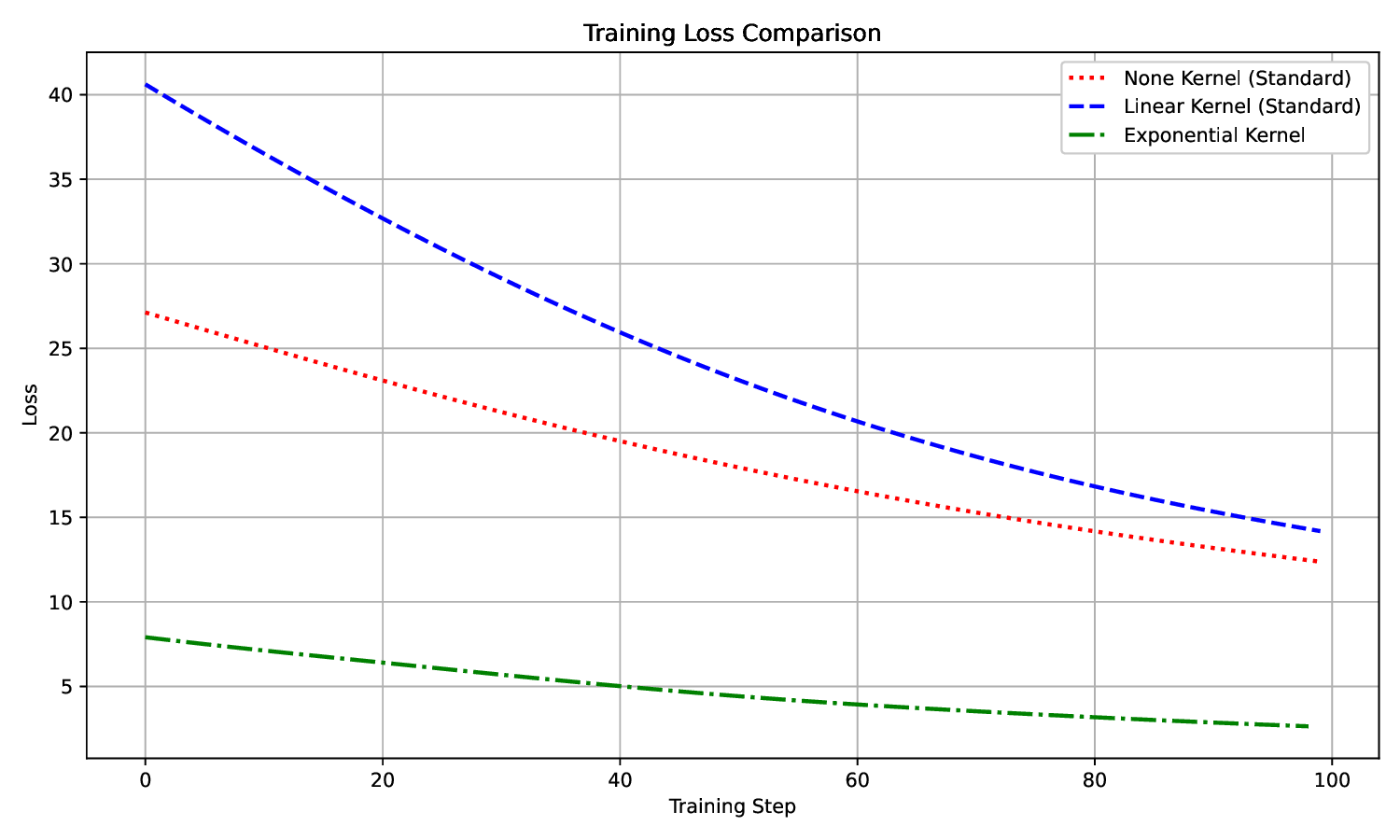}%
}
\caption{(\textbf{a}) Comparison of function approximations under different constraint embedding kernels: analytical solution (solid black line), no kernel embedding (red dashed line), linear kernel embedding (blue dashed line), and exponential kernel embedding (green dash-dotted line). (\textbf{b}) Evolution of training loss for models with different constraint embedding kernels.}
\label{fistpolycomp}
\end{figure}

Beyond the quantitative efficiency gains, the performance advantage of our approach is further illustrated in Fig.~\ref{fistpolycomp}. Fig.~\ref{fistpolycomp}(\textbf{a}) compares function approximations under different constraint embedding kernels. The analytical solution (solid black line) is compared against predictions from three quantum circuit models employing distinct hard-constraint embedding strategies: None kernel (red dashed line), Linear kernel (linear hard constraint, blue dashed line), and Exponential kernel (exponential hard constraint, green dash-dotted line). All models were trained under identical settings (number of qubits, ansatz depth, and optimization steps). The exponential kernel embedding tracks the analytical solution most closely, especially near the boundary region, confirming that the exponential kernel embedding effectively enforces the initial condition while maintaining the model's expressive power.

Fig.~\ref{fistpolycomp}(\textbf{b}) compares evolution of training loss for models with different constraint embedding kernels. Training loss is plotted as a function of optimization steps for the three kernel variants: None kernel (red dashed line), Linear kernel (blue dashed line), and Exponential kernel (green dash-dotted line). The exponential kernel embedding achieves the fastest convergence and the lowest final loss, outperforming both the linear kernel and the none kernel baseline. This indicates that the exponential hard constraint embedding not only improves solution accuracy but also stabilizes and accelerates the optimization process in physics-informed quantum machine learning. The None kernel is particularly inefficient because, without any constraint embedding, the initial condition is enforced solely through the loss function as a soft penalty. This forces the optimizer to simultaneously balance the boundary condition satisfaction with the differential equation residual, leading to a harder optimization landscape, slower convergence, and a higher final loss. In contrast, both the linear and exponential kernels eliminate this burden by analytically enforcing the initial condition, thereby reducing the optimization to fitting the differential equation residual alone.

\subsection{Parametrized Damped Oscillator Equation}

We further tested our method with a more representative physical problem described by the parametrized damped oscillator equation,
\begin{equation}
\frac{{dy}}{{dx}} + \lambda \kappa y + \lambda \tan (\lambda x)y = 0,\quad y(0) = 1,
\end{equation}
where $\lambda$ and $\kappa$ are real parameters.

This equation has an exact analytical solution of the form
\begin{equation}
y(x) = {e^{( - \kappa \lambda x)}}\cos (\lambda x) + \text{const},
\end{equation}
with the constant determined by the initial condition.

\begin{figure}[htb]
\centering
\subfloat[]{%
\includegraphics[width=0.48\linewidth]{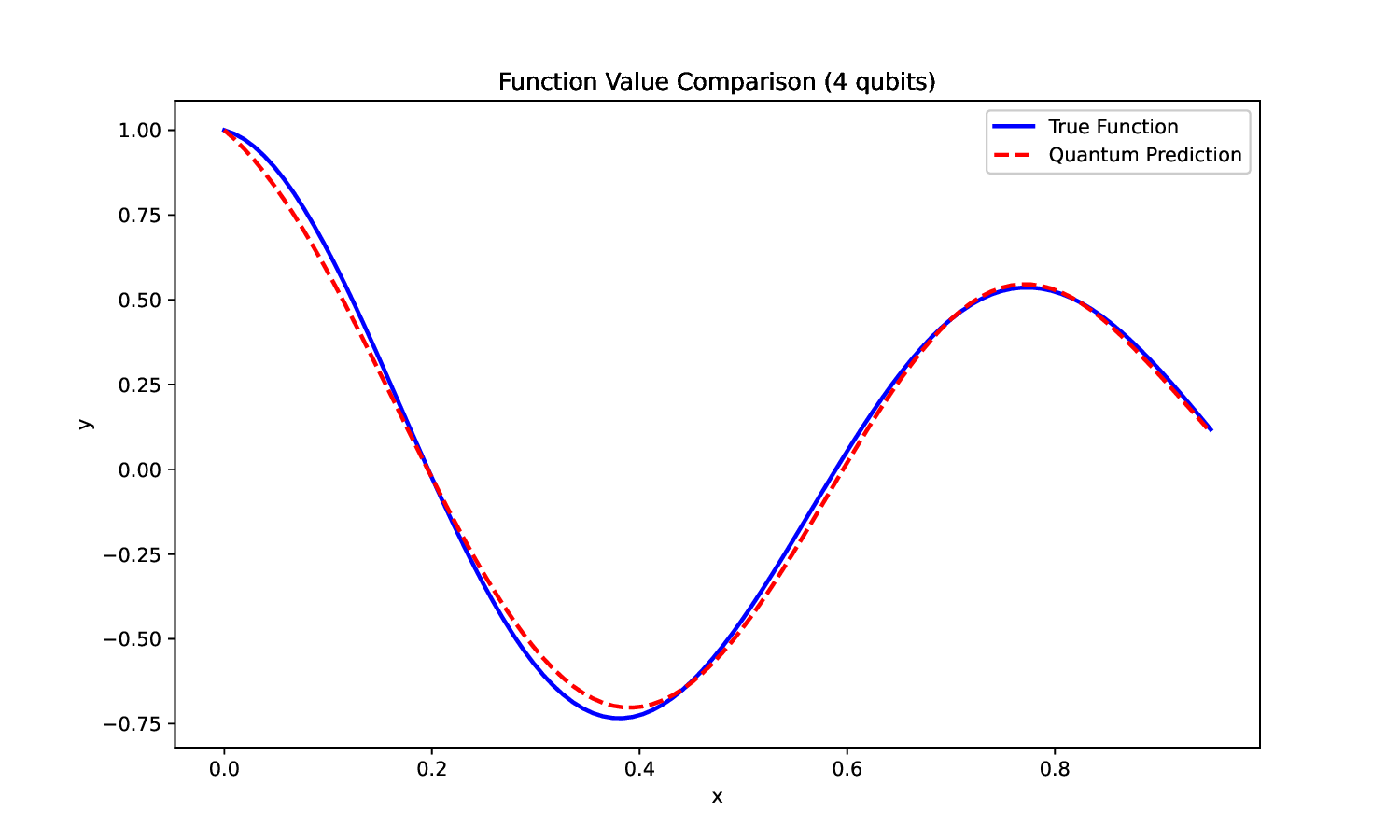}%
}%
\hfill%
\subfloat[]{%
\includegraphics[width=0.48\linewidth]{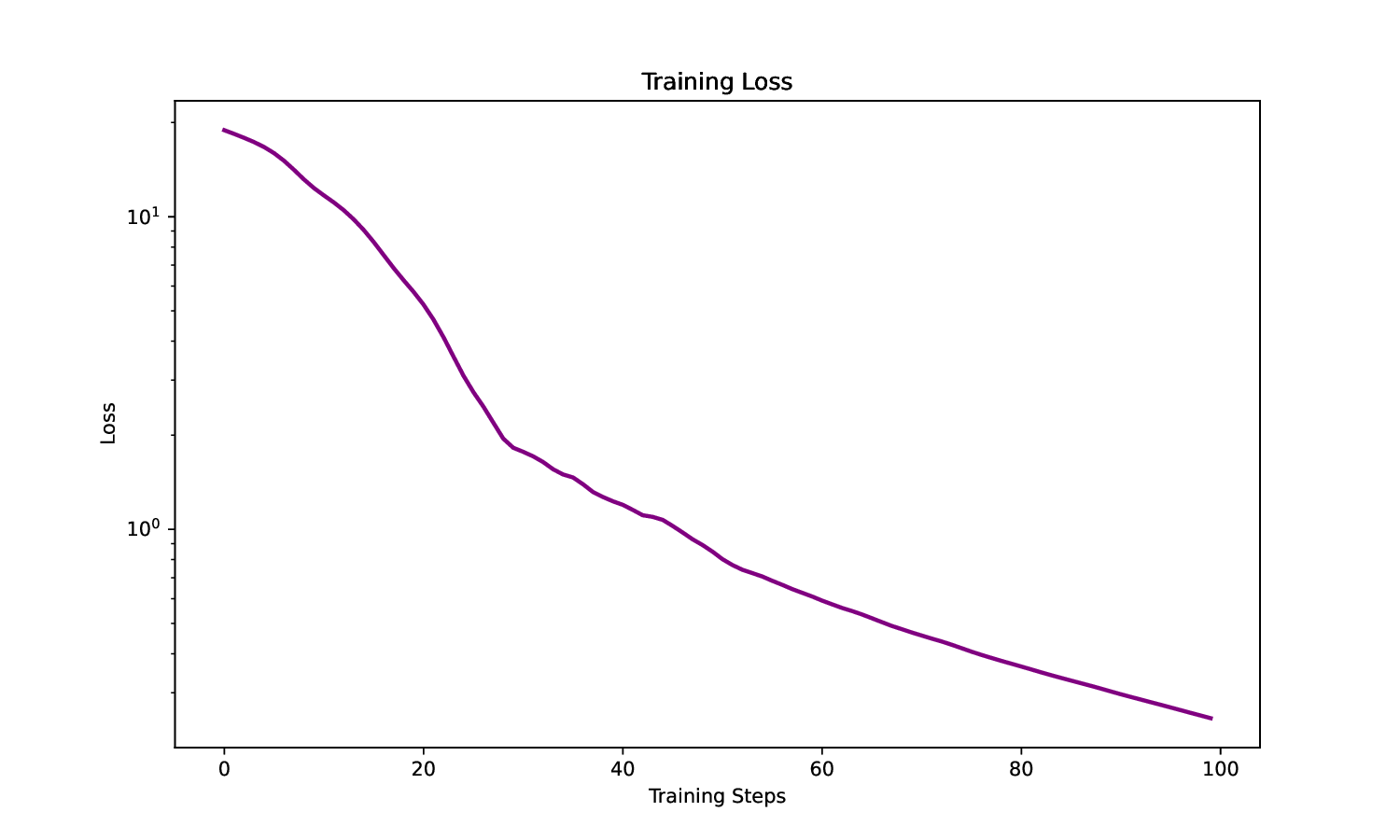}%
}
\caption{(\textbf{a}) Comparison of the true solution (blue solid line) and the PIQML prediction (red dashed line) for the parametrized damped oscillator equation with $\lambda = 8,\kappa = 0.1$ over the domain $[0,1]$. (\textbf{b}) Training loss as a function of optimization steps.}
\label{damped1}
\end{figure}

For numerical demonstration, we set $\lambda = 8,\kappa = 0.1$. The model is configured with 4 qubits, 4 layers of the variational ansatz and 32 parameters, and trained for 100 optimization steps. Fig.~\ref{damped1}(\textbf{a}) compares the true function (blue solid line) with the prediction from our PIQML model (red dashed line) over the domain interval $[0, 1]$. The quantum predictions closely match the true function over the entire interval, demonstrating the model's ability to accurately capture oscillatory and exponentially decaying behavior. Fig.~\ref{damped1}(\textbf{b}) shows the training loss as a function of optimization steps. The loss decreases monotonically and stabilizes at a low value, indicating stable convergence without noticeable oscillations or overfitting. Most of the error reduction occurs in the early stages of training, highlighting the efficiency of the optimization process. We note that extending the training beyond 100 steps could further reduce the final loss to values below $10^{-1}$, as the loss curve indicates continued potential for reduction.

\begin{figure}[htb]
\centering
\subfloat[]{%
\includegraphics[width=0.48\linewidth]{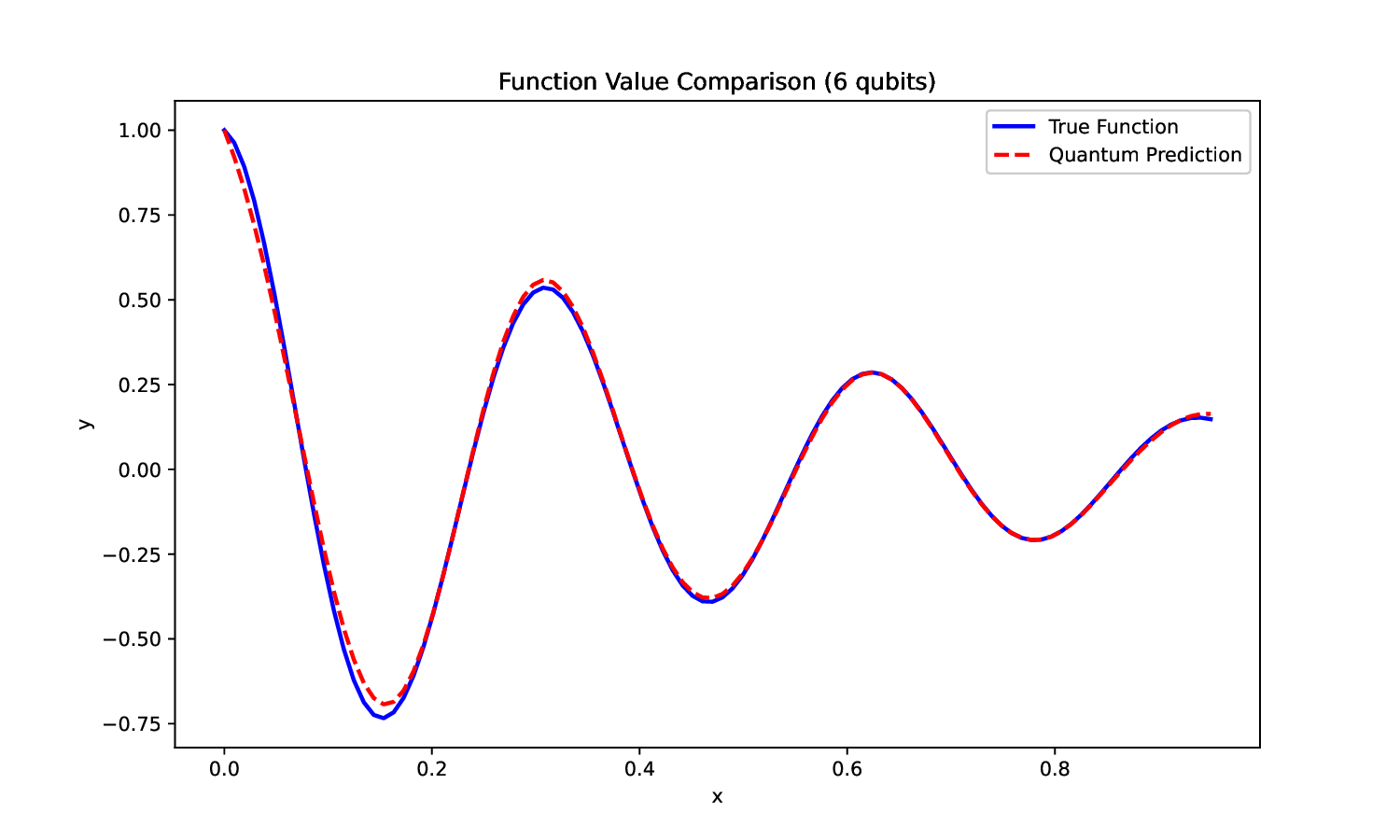}%
}%
\hfill%
\subfloat[]{%
\includegraphics[width=0.48\linewidth]{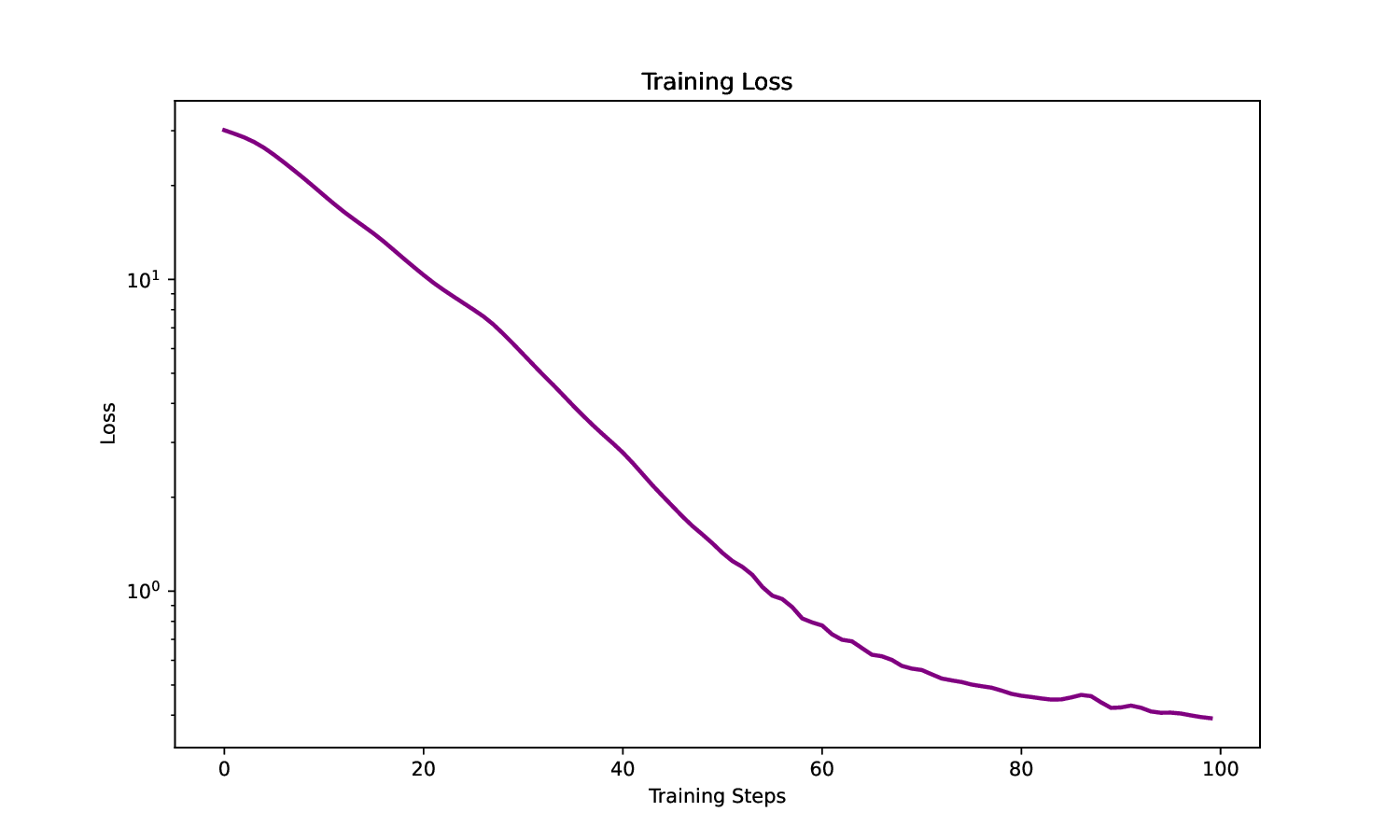}%
}
\caption{(\textbf{a}) Comparison of the true target function (blue solid line) and the PIQML prediction (red dashed line) for the parametrized damped oscillator equation with $\lambda = 20,\kappa = 0.1$ using 6 qubits, over the input range $[0,1]$. (\textbf{b}) Training loss as a function of optimization steps.}
\label{damped2}
\end{figure}

In the second case with $\lambda = 20,\kappa = 0.1$, the model is configured with 6 qubits, 4 layers of the variational ansatz and 48 parameters, and trained for 100 optimization steps. Fig.~\ref{damped2}(\textbf{a}) compares the true target function (blue solid line) with our PIQML model (red dashed line) for a system with 6 qubits. The true function exhibits a strongly oscillatory pattern due to the high frequency parameter $\lambda = 20$, while the quantum model successfully captures both the oscillatory behavior and the overall shape across the input range $[0,1]$. The amplitude of the oscillations remains consistent without significant decay, reflecting the minimal damping effect from $\kappa = 0.1$. Both curves align closely, indicating accurate learning and approximation capability of the quantum model under these parameters. In Fig.~\ref{damped2}(\textbf{b}), the loss decreases rapidly within the first 20 steps, dropping by an order of magnitude from above ${10}^1$ to around ${10}^0$, and then continues to decline gradually until reaching a stable low plateau. The smooth convergence without significant fluctuations suggests stable training dynamics, despite the increased model complexity from using 6 qubits and the high-frequency target function ($\lambda = 20$). The minimal effective damping ($\kappa = 0.1$) does not hinder the learning process, as evidenced by the consistent downward trend in loss.

\subsection{Nonlinear differential equations}

To evaluate the capability of our method for nonlinear problems, we solved the following nonlinear oscillatory equation,
\begin{equation}
\frac{{dy}}{{dx}} - 4y + 6{y^2} - \sin (50x) - y\cos (25x) + 1/2 = 0,\quad y(0) = 0.75
\end{equation}

The classical numerical reference solution is obtained using the Runge--Kutta method (specifically, the RK45 adaptive solver) implemented in the SciPy library, which serves as the ground truth for evaluating the quantum model's accuracy.

All numerical experiments presented in this work were carried out as classical simulations of quantum circuits. The simulations were implemented using the PennyLane and NumPy libraries in Python, running on classical hardware. No actual NISQ quantum hardware was used.

\begin{figure}[htb]
\centering
\includegraphics[width=\linewidth]{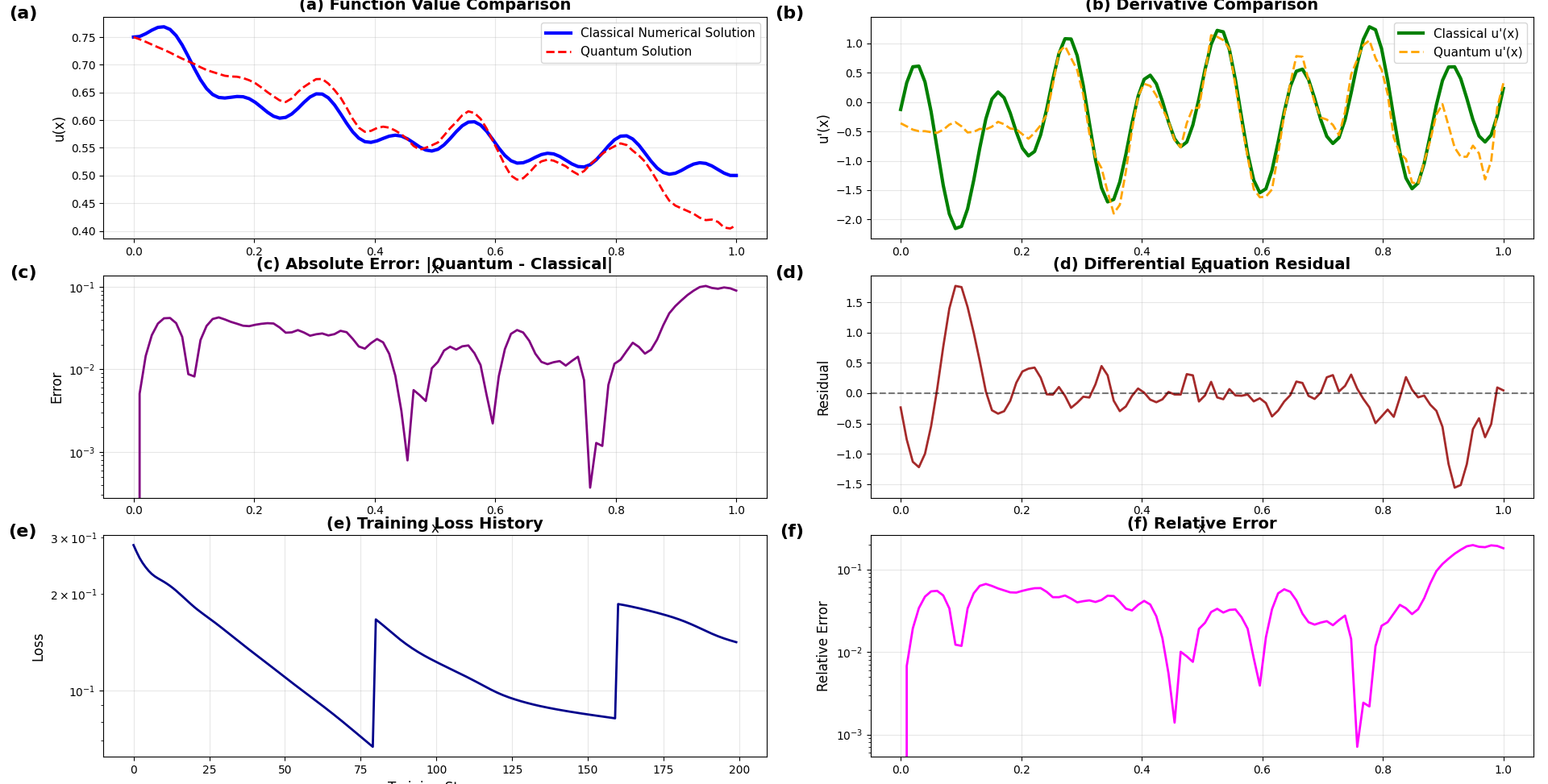}
\caption{Comprehensive evaluation of the proposed PIQML framework on a nonlinear oscillatory differential equation. Subfigures (\textbf{a}) and (\textbf{b}) compare the solution function and its derivative against a classical reference. Subfigures (\textbf{c}) and (\textbf{d}) quantify the absolute and relative errors, respectively. Subfigure (\textbf{e}) visualizes the residual, and (\textbf{f}) shows the convergence history of the training loss.}\label{node}
\end{figure}

Fig.~\ref{node} provides a comprehensive comparison of the results, with each subfigure visualizing a key aspect of the analysis:

(\textbf{a}) Function Value Comparison: This panel presents a direct comparison between the solution obtained from our PIQML method (red dashed line) and the classical numerical reference solution (blue solid line) across the domain. The close alignment between the two curves provides a visual confirmation of the PIQML model's accuracy in capturing the overall profile of the solution.

(\textbf{b}) Derivative Comparison: This figure presents a comparison between the derivative of the quantum simulation function (yellow dashed line) and the derivative of the classical function (green solid line). The consistency between the two, particularly in the central region of the domain, demonstrates that the quantum model successfully captures not only the function values but also the underlying dynamical behavior governed by the differential equation. Minor deviations observed near the boundaries are consistent with the error patterns seen in subfigure (\textbf{a}).

(\textbf{c}) Absolute Error: This map quantifies the pointwise absolute difference between the PIQML model and classical solutions $(\mid y_{\mathrm{Quantum}}-y_{\mathrm{Classical}}\mid)$. Regions with brighter colors indicate lower error magnitude.

(\textbf{d}) Differential Equation Residual: This plot visualizes the residual $\mathcal{L}_{\mathrm{res}}^{\left(i\right)}\left(\mathbf{\Theta}\right)=\left[\frac{dy(x_i)}{dx}-g\left(x_i,y(x)\right)\right]^2$. It measures the extent to which the PIQML solution satisfies the original differential equation at every point in the domain. A uniformly low residual across the domain is a strong indicator that the physical laws have been effectively embedded into the learned model.

(\textbf{e}) Training Loss History: This curve tracks the total loss $\mathcal{L}_{\mathrm{total}}(\mathbf{\Theta}) = \frac{1}{M} \sum_{i=1}^{M} \left[ L_{\mathrm{res}}^{(i)}(\mathbf{\Theta}) + L_{\mathrm{data}}^{(i)}(\mathbf{\Theta}) \right]$ against the number of training iterations. A monotonically decreasing trend that converges to a low plateau demonstrates the stability and effectiveness of the training process. The convergence rate and final loss value are key metrics for evaluating training performance.

(\textbf{f}) Relative Error: This plot shows the pointwise relative error, calculated as $(\mid y_{\mathrm{quantum}}-u_{\mathrm{classical}}\mid/\mid u_{\mathrm{classical}}\mid)$. Normalizing the error by the magnitude of the true solution provides insight into the significance of the discrepancies, especially in regions where the solution value is small and absolute error might be misleading.

\section{Conclusion}\label{sec4}

This paper presented and numerically validated a novel physics-informed quantum machine learning framework with hard constraint embedding, specifically designed for the NISQ era, for solving nonlinear differential equations.

The core contributions of this work are threefold. First, we designed a rigorous hard constraint embedding mechanism that analytically and exactly enforces initial/boundary conditions into the trial solution, eliminating the constraint violation errors and optimization balancing challenges inherent in traditional soft-constraint methods. Second, we implemented a fully quantum-native gradient computation, utilizing the parameter-shift rule to calculate derivatives with respect to the input variable directly, bypassing the classical bottleneck of discretizing the equation into a linear system and leveraging a unique property of quantum computation. Finally, we integrated NISQ-compatible variational quantum circuits with tunable feature maps (e.g., Chebyshev and Fourier maps), enabling the model to capture a wide range of complex behaviors, from smooth variations to high-frequency oscillations, using shallow circuits.

Comprehensive numerical experiments confirmed the superiority of the proposed framework. On multiple benchmark problems---linear, nonlinear, and damped oscillatory---our method achieved high-accuracy solutions using fewer qubits, shallower circuit depths, and fewer optimization steps than existing approaches. The introduction of the Fourier feature map, in particular, dramatically improved the model's capability to handle high-frequency problems.

\textbf{Quantum advantage outlook.} The central question of whether and when the proposed quantum framework can outperform classical methods remains open. No complexity-theoretic analysis or scaling study is provided in this work to establish a quantum advantage. The potential for quantum advantage in variational quantum machine learning for differential equations may arise from the exponentially large Hilbert space accessible with polynomially many qubits, which could, in principle, provide a more compact representation of complex functions than classical neural networks. However, realizing this potential advantage requires overcoming significant challenges, including barren plateaus in training, noise on NISQ hardware, and the overhead of parameter-shift gradient evaluation. Rigorous theoretical analysis and large-scale empirical studies are needed to determine the regimes where quantum methods may offer genuine advantages over classical alternatives.

Looking forward, this framework paves the way for solving complex partial differential equations from science and engineering on near-term quantum devices. Future work will focus on extending the method to higher-dimensional PDEs, exploring more efficient strategies for trainable frequency selection, and ultimately demonstrating its performance on physical quantum hardware. We believe that the deep integration of physical priors with the expressive power of quantum models through hard constraints is a key pathway toward realizing practical quantum advantage.


\section*{Acknowledgments}
The research was funded by the Natural Science Foundation of Zhejiang Province, China (Grant No.~Q24A050004).

\section*{Conflicts of Interest}
The authors declare no conflicts of interest.


\bibliographystyle{unsrtnat}
\bibliography{ref}

\end{document}